\documentclass[10pt,conference]{IEEEtran}
\IEEEoverridecommandlockouts
\usepackage{cite}
\usepackage{amsmath,amssymb,amsfonts}
\usepackage{graphicx}
\usepackage{textcomp}
\usepackage{xcolor}
\usepackage[hyphens]{url}
\usepackage{fancyhdr}
\usepackage{hyperref}

\usepackage[ruled,linesnumbered]{algorithm2e}

\SetCommentSty{mycommfont}

\usepackage{comment}
\usepackage[numbers]{natbib}

\usepackage{braket}
\usepackage{float}
\usepackage{multirow}
\usepackage{makecell}
\usepackage{bm}
\usepackage{subfig}
\usepackage{setspace}
\usepackage{booktabs}

\def\BibTeX{{\rm B\kern-.05em{\sc i\kern-.025em b}\kern-.08em
    T\kern-.1667em\lower.7ex\hbox{E}\kern-.125emX}}

\title{
QueenV2: Future of Quantum Circuit Simulation\\
{
\footnotesize{Seven Seas of Rhye}
}
}

\author{
    \IEEEauthorblockN{Chuan-Chi Wang}
    \IEEEauthorblockA{National Taiwan University, Taipei, Taiwan \\
    Email: d10922012@ntu.edu.tw}
}

\begin{document}
\maketitle

\ifdefined\hpcacameraready 
  \thispagestyle{camerareadyfirstpage}
  \pagestyle{empty}
\else
  \thispagestyle{plain}
  \pagestyle{plain}
\fi

\newcommand{\hpcaheight}{0mm}
\ifdefined\eaopen
\renewcommand{\hpcaheight}{12mm}
\fi

\begin{abstract}

A state vector-based quantum circuit simulation can provide accurate results for the development and validation of quantum computing algorithms, without being affected by noise interference.
However, existing quantum circuit simulators have consistently underperformed due to inadequate integration with quantum circuits and high-performance computing architectures.
To tackle the challenges in quantum computing, we propose QueenV2, which builds upon the design principles of Queen and elevates performance to a new level.
Experimental results on the NVIDIA RTX-4090 demonstrate that QueenV2 achieves up to a 40x improvement in gate performance and a 5x improvement in circuit performance compared to hyQuas.
Furthermore, QueenV2 realizes a 137x speedup in gate benchmarks and a 14x speedup in circuit performance relative to NVIDIA cuQuantum, enabled by gate fusion via the IBM Qiskit toolkit.
By eliminating reliance on third-party libraries, ~\emph{QueenV2} is positioned to significantly accelerate quantum circuit simulation, thus promoting the development of innovative accelerators and quantum algorithms.

\end{abstract}

\section{Introduction}
\label{sec:introduction}
Please refer to the introduction of Queen~\cite{Queen}.
The main issue is that previous simulators were too slow, so more comprehensive optimizations were necessary.
This paper complements a direct comparison with hyQuas~\cite{HyQuas}, which uses the hybrid technique to improve the data locality.
Moreover, the additional techniques and illustrations are also revealed.


\section{Background}
\label{sec:background}
In this section, we introduce the fundamental quantum gates utilized in QCS as our previous work, Queen~\cite{Queen}, along with the mechanisms for high-performance computing for QCS.

\subsection{Qubit Representation and Quantum Gates}
\label{sec:quantum_gate}
A qubit is the fundamental unit of computation in quantum computing, characterized by the state $\alpha\ket{0} + \beta\ket{1}$, where $\ket{0}=[1,0]^T$ and $\ket{1}=[0,1]^T$ in Dirac notation.
The non-deterministic behaviour of a qubit, known as superposition, allows it to exist in both states $\ket{0}$ and $\ket{1}$ simultaneously with probabilities $|\alpha|^2$ and $|\beta|^2$, respectively.
It is imperative to ensure that the probabilities must satisfy the normalization condition $|\alpha|^2 + |\beta|^2 = 1$ for a single-qubit system.

When extending this concept to multiple qubits, the quantum state becomes a superposition of $2^\mathit{N}$ basis state, ranging from $\ket{0}$ to $\ket{1}$. The $\mathit{N}$-qubit quantum state can be formulated as Equation~\ref{eq:state}, where $a_i^2$ describes the probability of the $\ket{i}$.

\begin{equation}
\ket{\phi} = \sum_{i \in [0, 2^N)} a_i \ket{i}
\label{eq:state}
\end{equation}

Quantum circuits comprise various quantum logic gates, which are the basic units of quantum operation. They can be typically categorized into two main types: one-qubit gates and multi-qubit gates.
Consider a one-qubit gate; the matrix representation for the state vector is provided by Equation~\ref{eq:one_qubit_gate}, where the 2x2 matrix $G$ represents a one-qubit gate operating on a specific qubit and the $0_{i}$ and $1_{i}$ denote the amplitudes that the $i$-th bit in bitstring is 0 or 1.

\begin{equation}
\begin{bmatrix}
a^{\prime}_{*...*0_{i}*...*} \\
a^{\prime}_{*...*1_{i}*...*}
\end{bmatrix}
\mapsto G
\begin{bmatrix}
a^{}_{*...*0_{i}*...*} \\
a^{}_{*...*1_{i}*...*}
\end{bmatrix}
\label{eq:one_qubit_gate}
\end{equation}

The simulation of an $N$-qubit gate, represented by a $2^N \times 2^N$ matrix, involves a similar process.
For instance, consider a 2-qubit CNOT gate acting on the $i$-th and $j$-th qubits as a typical example.
The processes can be expressed as Equation~\ref{eq:cnot}.

\begin{equation}
\begin{bmatrix}
a^{\prime}_{*...*0_{i}*..*0_{j}*...*} \\
a^{\prime}_{*...*0_{i}*..*1_{j}*...*} \\
a^{\prime}_{*...*1_{i}*..*0_{j}*...*} \\
a^{\prime}_{*...*1_{i}*..*1_{j}*...*}
\end{bmatrix}
\mapsto
\begin{bmatrix}
1 & 0 & 0 & 0 \\
0 & 1 & 0 & 0 \\
0 & 0 & 0 & 1 \\
0 & 0 & 1 & 0
\end{bmatrix}
\begin{bmatrix}
a_{*...*0_{i}*..*0_{j}*...*} \\
a_{*...*0_{i}*..*1_{j}*...*}  \\
a_{*...*1_{i}*..*0_{j}*...*} \\
a_{*...*1_{i}*..*1_{j}*...*}
\end{bmatrix}
\label{eq:cnot}
\end{equation}

\subsection{Optimized Techniques for High-Performance Computing}
\label{sec:opt4hpc}
To achieve effective state vector-based QCS, it is essential to incorporate complementary optimization techniques such as optimizing data locality, minimizing unnecessary operations, and rearranging the order of gates.
The most effective mechanisms known to date include but are not limited to, \emph{gate fusion}, \emph{qubit reordering}, and \emph{in-cache operations}.

\emph{\textbf{Gate Fusion.}}
This technique is employed to merge multiple gates into a single generic gate~\cite{45qubit, qHiPSTER, HyQuas, Aer_sim}.
As the primary objective is to reduce the total number of quantum gates, this reduction can significantly decrease memory access in memory-bound simulators~\cite{cuQuantum, Aer_sim, QuEST, mpiQulacs}.
Despite these advantages, excessive gate fusion can not only result in longer quantum circuit transpile time but also lead to slower performance.
This issue arises because the index accessing pattern of fused quantum gates conflicts with the design principles of threads within a thread block in GPU architecture.

In most cases, a cost function is used to determine whether merging gates is necessary, and optimal performance is typically achieved when merging 3 to 5 qubits.
Among various fusion techniques, diagonal optimization is a proven method that can simultaneously reduce the number of gates while maintaining computational efficiency~\cite{qaoa_arxiv}.

\emph{\textbf{Qubit Reordering.}}
The deliberate integration of supplementary swaps within the circuit indicates the requirement to adjust the indexing of subsequent gates.
Although this approach needs extra operations, the qubits for the subsequent gates have already been rearranged to the least significant bits for the classical computers.
This signifies that the state represented by that qubit has been positioned closer to the computational unit, resulting in substantial improvement in data locality.
In modern QCS, the qubit swapping optimization~\cite{mpiQulacs, cache_blocking, cpu_gpu_communication} is incorporated to reduce communication overhead across ranks\footnote{Given its common usage in data transfer APIs, we use the term \emph{ranks} to encompass machines, nodes, and devices collectively.}.

To the best of our knowledge, Queen~\cite{Queen} is the first work to utilize a comprehensive qubit reordering mechanism in a single swapping operation to achieve outstanding performance and optimize data locality in both single-rank and multi-rank scenarios.
In a single-rank environment, this approach significantly reduces the number of caching misses by facilitating continuous gate operations in the gate block. In a multi-rank setup, it ensures data continuity, substantially maximizing the transfer bandwidth and minimizing the total transfer time.

\emph{\textbf{In-cache Operations.}}
In most modern quantum circuit simulators such as cuQuantum~\cite{cuQuantum}, IBM Aer Simulator~\cite{Aer_sim}, QuEST~\cite{QuEST}, and Google Cirq, gate operations are computed using matrix multiplications and simulated in a gate-by-gate scheme.
However, those legacy approaches fail to utilize the computational mechanisms of classical computers fully, thereby impeding the adoption of advanced high-performance techniques.
To rectify the situation, Queen~\cite{Queen} introduces the all-in-cache simulation module with block-by-block and an all-in-one optimization module to expedite simulation and eliminate performance bottlenecks.

While HyQuas~\cite{HyQuas} and UniQ~\cite{uniq} do not fully capitalize on the caching potential for all quantum gates with their hybrid strategies, high-performance libraries such as cuTT~\cite{cutt} and cuBLAS~\cite{cublas} endowed them with a substantial performance advantage over other memory-bound of QCS frameworks.


\section{QueenV2}
\label{sec:queenv2}
\begin{figure}[tb!]
\centerline{\includegraphics[width=.95\columnwidth]{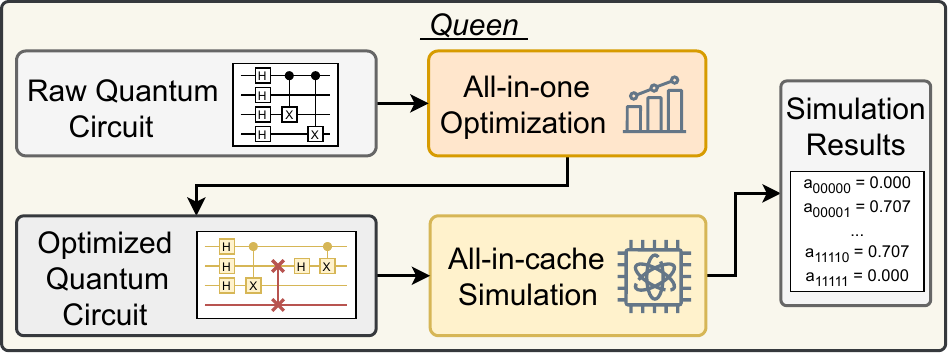}}
\caption{QueenV2 workflow for quantum simulation.}
\label{fig:workflow}
\end{figure}

As illustrated in Figure~\ref{fig:workflow}, the QueenV2 workflow fully inherits from the first generation while incorporating significant modifications in both the all-in-one optimization module (AIO) and the all-in-cache simulation module (AIC).
As the author is currently learning about other research topics, a quick overview of the techniques is provided for peers and researchers.
Once this phase is completed, any implementation details and knowledge corrections will be shared soon.

\subsection{All-in-one circuit module}
\label{sec:aio}
The AIO module needs to be elevated from a three-tier architecture to a five-tier architecture.
The two additional stages required are the tensor product booster and infinite diagonal gate fusion~\cite{qaoa_arxiv}.

The overall computational flow is as follows:
\begin{enumerate}
    \item Find \emph{\textbf{Tensor Product Booster}} of Gate Blocks.
    \item Find \emph{\textbf{Infinite diagonal Gate fusion}} of Gate Blocks.
    \item Find \emph{\textbf{Cross-Rank}} of Gate Blocks.
    \item Find \emph{\textbf{In-Rank}} of Gate Blocks.
    \item Find \emph{\textbf{General Gate fusion}} of Gate Blocks.
\end{enumerate}

The module utilizes the upgraded version of the findGB algorithm from Queen~\cite{Queen}, considering the control qubits. In the case of infinite diagonal gate fusion, additional gate reordering is essential to ensure that all diagonal gates are visible for subsequent operations.
The concept of the \emph{\textbf{Tensor Product Booster}} is inspired by merge sort for performance improvement; it will be shared in more detail with everyone in the future.

\begin{figure*}[tb!]
\centerline{\includegraphics[width=1.98\columnwidth]{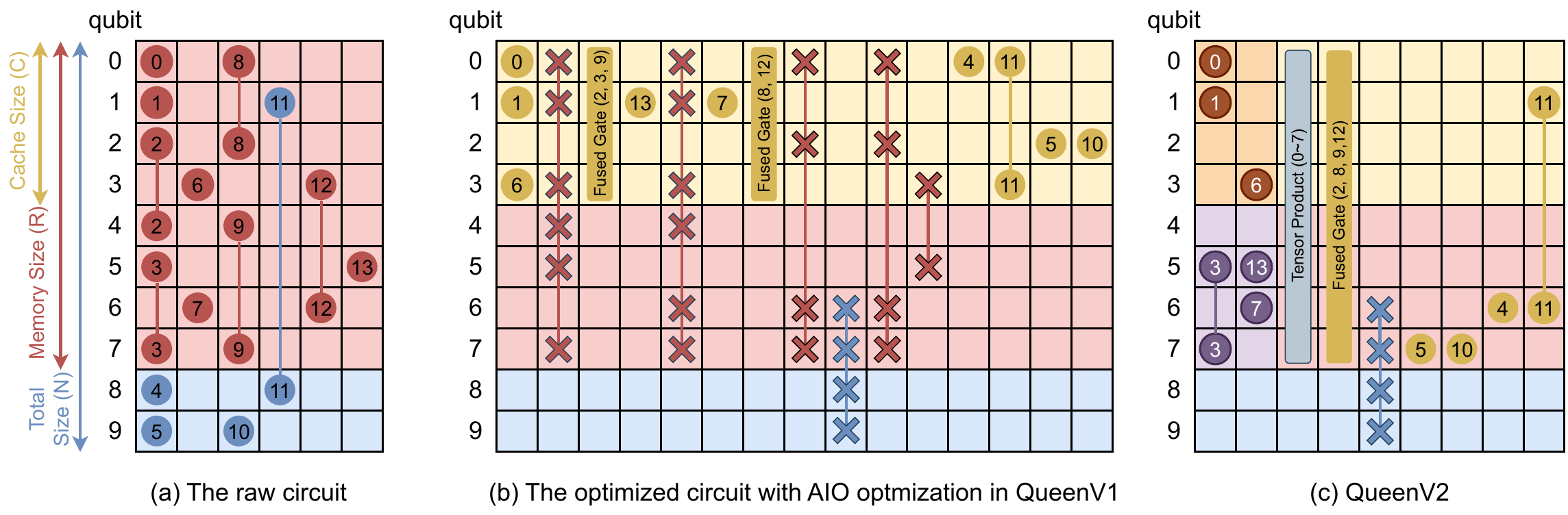}}
\caption{The circuit diagram for the simulator of QueenV1 and QueenV2.}
\label{fig:queenV2}
\end{figure*}

The overall schematic is illustrated in Figure~\ref{fig:queenV2}.
To ensure that everyone can implement the qubit reordering mechanism, please prepare a bi-directly corresponding hash table for all qubits.
The orange and purple regions represent the gate block sections for Tensor Product Booster (TPB).
The diagonal gates will be merged as much as possible when the conditions are met, and the 11th gate needs to be excluded due to it spanning across ranks.
The 3rd gate is not within the search scope because it is located in the TPB\#1 section.
Moreover, it is evident from the diagram that in-memory swapping (IMS) is no longer necessary in current QueenV2, with its use limited to instances of cross-rank swapping (CRS).

\subsection{All-in-cache simulation module}
\label{sec:aio}
This implementation builds upon the exemplary traditions of QueenV1 while minimizing dependency on third-party APIs.

The mechanism for reading from and writing back device memory has been incorporated through the implementation of instructions similar to the PEDP and PEXT instruction sets\footnote{If NVIDIA implements similar acceleration instructions, this part could benefit from further optimization.}.
Special thanks to Wang, my dear schoolmate, for his contributions in this area.
In CRS operations, IMS operations are still utilized due to the necessity of arranging sub-state vectors contiguously, facilitating efficient data transmission across multiple ranks to maximize the data transfer throughput.

The implementation of the Tensor Product Booster is based on the original concept of tensor products. This component was developed by my dear schoolmates, Yang and Chiu. The addition of cache-line alignment mechanisms can ensure the efficient operation of this functionality.

As with other incremental works, to support a wider range of special quantum gate operations in the simulator, one can initially generate them using the unitary gates currently supported by Queen.




\section{Evaluation}
\label{sec:evaluation}
Section~\ref{sec:env_setup} introduces the experimental environment and benchmark setup.
Subsequently, the performance results of the gate and circuit benchmarks are presented in Section~\ref{sec:gate_benchmark} and Section~\ref{sec:cir_benchmark}, respectively.
For details on scalability across multiple ranks, please refer to the results of Queen~\cite{Queen}, where strong scaling demonstrates that doubling the number of machines typically yields a twofold improvement.
Therefore, it will not be elaborated on here.

\subsection{Experimental Setup}
\label{sec:env_setup}
The hardware and software configurations are listed in \tablename~\ref{tab:platform}.
The CPU (AMD Ryzen 9 5950X) and the GPU (NVIDIA RTX 4090) are used for the quantum circuit simulation.
The operating system is Ubuntu 22.04 with kernel version 5.15.0-1044-nvidia, alongside CUDA toolkit version 12.3.107.

To ensure consistency and fairness in evaluating simulation performance, the average execution time is derived from 10 runs, adopting \emph{double}-precision floating-point numbers.
The experiments employ a 30-qubit configuration, which represents the maximum qubit capacity of an RTX-4090 GPU, covering the majority of performance results.
comprehensive and up-to-date comparisons, we utilize hyQuas\footnote{Due to the need to install the additional external libraries for executing UniQ and the minimal performance difference between UniQ and hyQuas on the GPU, hyQuas is presented as a representative case study here.}~\cite{HyQuas} and cuQuantum~\cite{cuQuantum} with the latest versions.
With similarly powerful gate fusion techniques as the IBM-Qiskit toolkit, these two works are the most representative examples of compute-bound and memory-bound simulators, respectively.

\begin{table} [tb!]
\caption{The hardware and software platforms.} 
  \label{tab:platform}
  \small{
  \begin{tabular}{p{1cm}p{6.8cm}}
    \toprule
   \textbf{Name} & \textbf{Attribute} \\
    \midrule
    CPU & AMD Ryzen 9 5950X Processor\\
    GPU & NVIDIA GeForce RTX 4090\\
    RAM & Kingston 128 GB DDR4 2400MHz \\ 
    OS  & Ubuntu 22.04 LTS (kernel version 5.19.0-43-generic)\\
    CUDA & CUDA Toolkit version 12.3.107 \\
    \bottomrule
\end{tabular}
  } 
\end{table}

\subsection{Gate Benchmark}
\label{sec:gate_benchmark}
Table~\ref{tab:h} presents the performance results of the Hadamard gate.
QueenV2 significantly outperforms both simulators, achieving more than 137x times the performance of NVIDIA cuQuantum in the 30-qubit case.
Conversely, hyQuas encounters out-of-memory conditions due to potential technical flaws.
Additionally, when the number of qubits is between 23 and 27, the performance of hyQuas demonstrates unexpectedly slow, indicating a need to revisit the overhead introduced by the cost function of their gate fusion module.

\begin{table}[bt!]
    \caption{The elapsed time of the Hadamard gate on the RTX-4090 (unit: seconds) is provided, with performance speedups labelled as SU followed by the respective values. The term OOM indicates out-of-memory conditions.}
    \label{tab:h}
    \setlength{\tabcolsep}{3.2mm}{
    \begin{tabular}{crrrrrr}
    \hline \specialrule{0em}{1.5pt}{1.5pt}
    Qubit & QueenV2 & hyQuas (SU) & cuQuantum (SU) \\
    \hline
23 & 0.011 & 1.168 (\textbf{99.8x}) & 0.071	(\textbf{6.1x}) \\
24 & 0.014 & 1.297 (\textbf{90.7x}) & 0.165	(\textbf{11.6x}) \\
25 & 0.014 & 0.999 (\textbf{67.5x}) & 0.320	(\textbf{21.7x}) \\
26 & 0.021 & 1.305 (\textbf{60.1x}) & 0.641	(\textbf{29.6x}) \\
27 & 0.021 & 1.440 (\textbf{67.0x}) & 1.311	(\textbf{61.0x}) \\
28 & 0.035 & 1.557 (\textbf{44.1x}) & 2.704	(\textbf{76.6x}) \\
29 & 0.041 & 1.655 (\textbf{44.0x}) & 5.580	(\textbf{135.1x})\\
30 & 0.084 & OOM (\textbf{$\infty$})  & 11.572 (\textbf{137.4x}) \\
    \hline
    \end{tabular}
    }
\end{table}

\subsection{Circuit Benchmark}
\label{sec:cir_benchmark}
In the circuit benchmark, we employed the common circuit, including the Quantum Fourier Transform (QFT) and the Quantum Approximate Optimization Algorithm (QAOA) with 5 layers and full connectivity.
Since hyQuas forces to enable gate fusion, we will not present values for Queen and cuQuantum without fusion in these comparisons.

Table~\ref{tab:qft} presents the performance results of the QFT.
QueenV2 continues to outperform other simulators, achieving up to 15.5x times the performance of cuQuantum.
Although hyQuas still encounters out-of-memory conditions, QueenV2 is only up to 1.9x times faster. This indicates that hyQuas is indeed a compute-bound simulator with optimizations for control qubits, as mentioned in their paper.

\begin{table}[bt!]
    \caption{The elapsed time of QFT (unit: seconds).}
    \label{tab:qft}
    \setlength{\tabcolsep}{3.2mm}{
    \begin{tabular}{crrrrrr}
    \hline \specialrule{0em}{1.5pt}{1.5pt}
    Qubit & QueenV2 & hyQuas (SU) & cuQuantum (SU) \\
    \hline
23 & 0.011 & 0.012 (\textbf{1.0x}) & 0.0224	(\textbf{2.0x}) \\
24 & 0.014 & 0.017 (\textbf{1.2x}) & 0.1462	(\textbf{9.8x}) \\
25 & 0.024 & 0.031 (\textbf{1.3x}) & 0.2461	(\textbf{10.1x}) \\
26 & 0.041 & 0.058 (\textbf{1.4x}) & 0.459	(\textbf{11.1x}) \\
27 & 0.064 & 0.113 (\textbf{1.8x}) & 0.9055	(\textbf{14.0x}) \\
28 & 0.124 & 0.227 (\textbf{1.8x}) & 1.8442	(\textbf{14.8x}) \\
29 & 0.252 & 0.479 (\textbf{1.9x}) & 3.8149	(\textbf{15.1x})\\
30 & 0.519 & OOM (\textbf{$\infty$})  & 8.0552 (\textbf{15.5x}) \\
    \hline
    \end{tabular}
    }
\end{table}

For the QAOA circuit, the fully connected nature signifies a more complex circuit, which allows for a greater demonstration of circuit optimization effectiveness.
As shown in Table~\ref{tab:qaoa}, QueenV2 achieves over 4.9x times the performance improvement over hyQuas and a 14-fold increase in performance compared to cuQuantum.

\begin{table}[bt!]
    \caption{The elapsed time of 5-level fully-connected QAOA (unit: seconds).}
    \label{tab:qaoa}
    \setlength{\tabcolsep}{3.2mm}{
    \begin{tabular}{crrrrrr}
    \hline \specialrule{0em}{1.5pt}{1.5pt}
    Qubit & QueenV2 & hyQuas (SU) & cuQuantum (SU) \\
    \hline
23 & 0.030 & 0.090 (\textbf{3.0x}) & 0.413 (\textbf{13.6x}) \\
24 & 0.054 & 0.187 (\textbf{3.5x}) & 0.519 (\textbf{9.6x}) \\
25 & 0.098 & 0.380 (\textbf{3.9x}) & 1.058 (\textbf{10.7x}) \\
26 & 0.196 & 0.811 (\textbf{4.1x}) & 2.201 (\textbf{11.2x}) \\
27 & 0.371 & 1.737 (\textbf{4.7x}) & 4.820 (\textbf{13.0x}) \\
28 & 0.776 & 3.717 (\textbf{4.8x}) & 10.413 (\textbf{13.4x}) \\
29 & 1.613 & 7.948 (\textbf{4.9x}) & 22.887 (\textbf{14.2x})\\
30 & 3.391 & OOM (\textbf{$\infty$})  & 47.590 (\textbf{14.0x}) \\
    \hline
    \end{tabular}
    }
\end{table}

\begin{figure}[tb!]
\centerline{\includegraphics[width=.95\columnwidth]{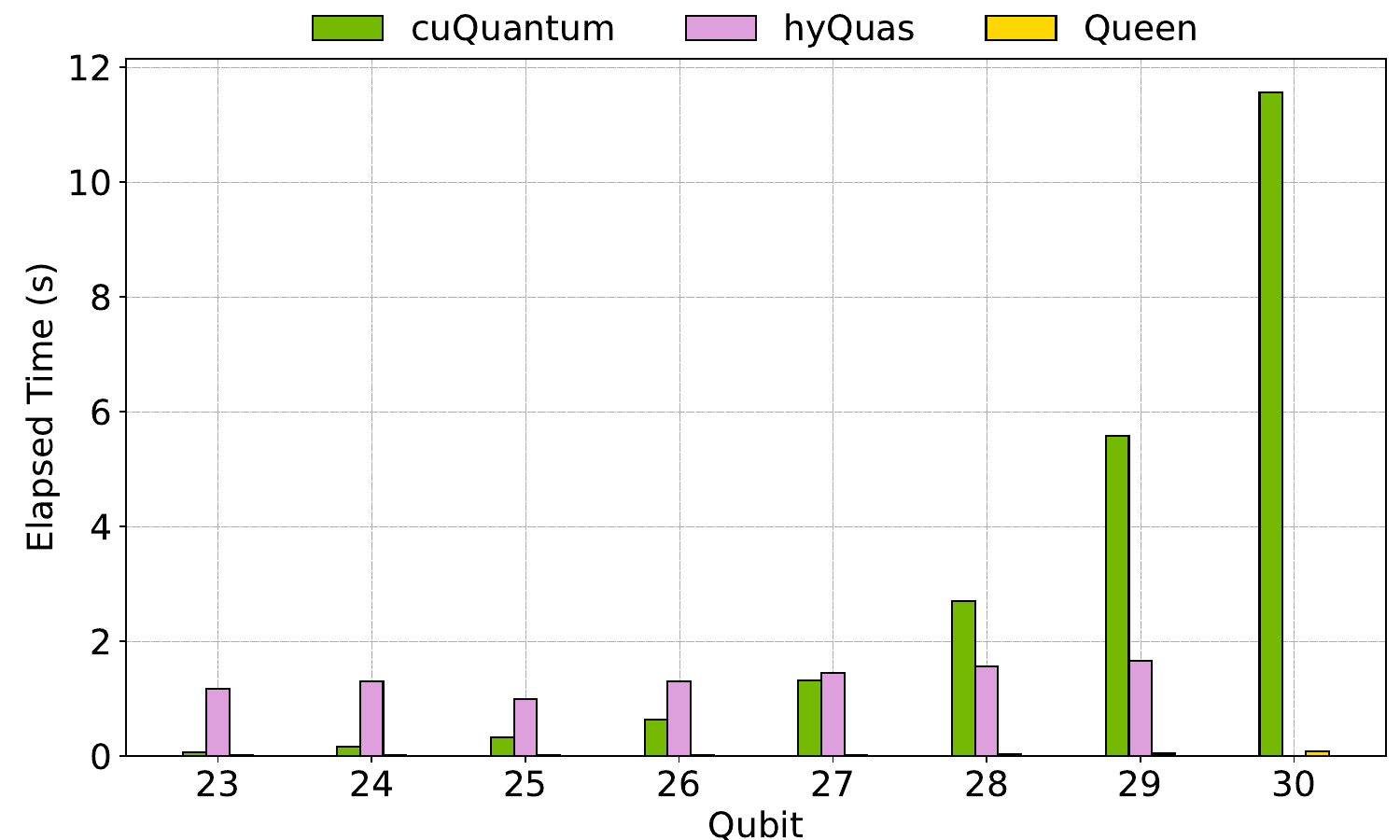}}
\caption{The H-gate benchmarks on RTX-4090 (The time of 0 seconds for hyQuas indicates out-of-memory).}
\label{fig:gate_benchmark}
\end{figure}

\begin{figure}[tb!]
\centerline{\includegraphics[width=.95\columnwidth]{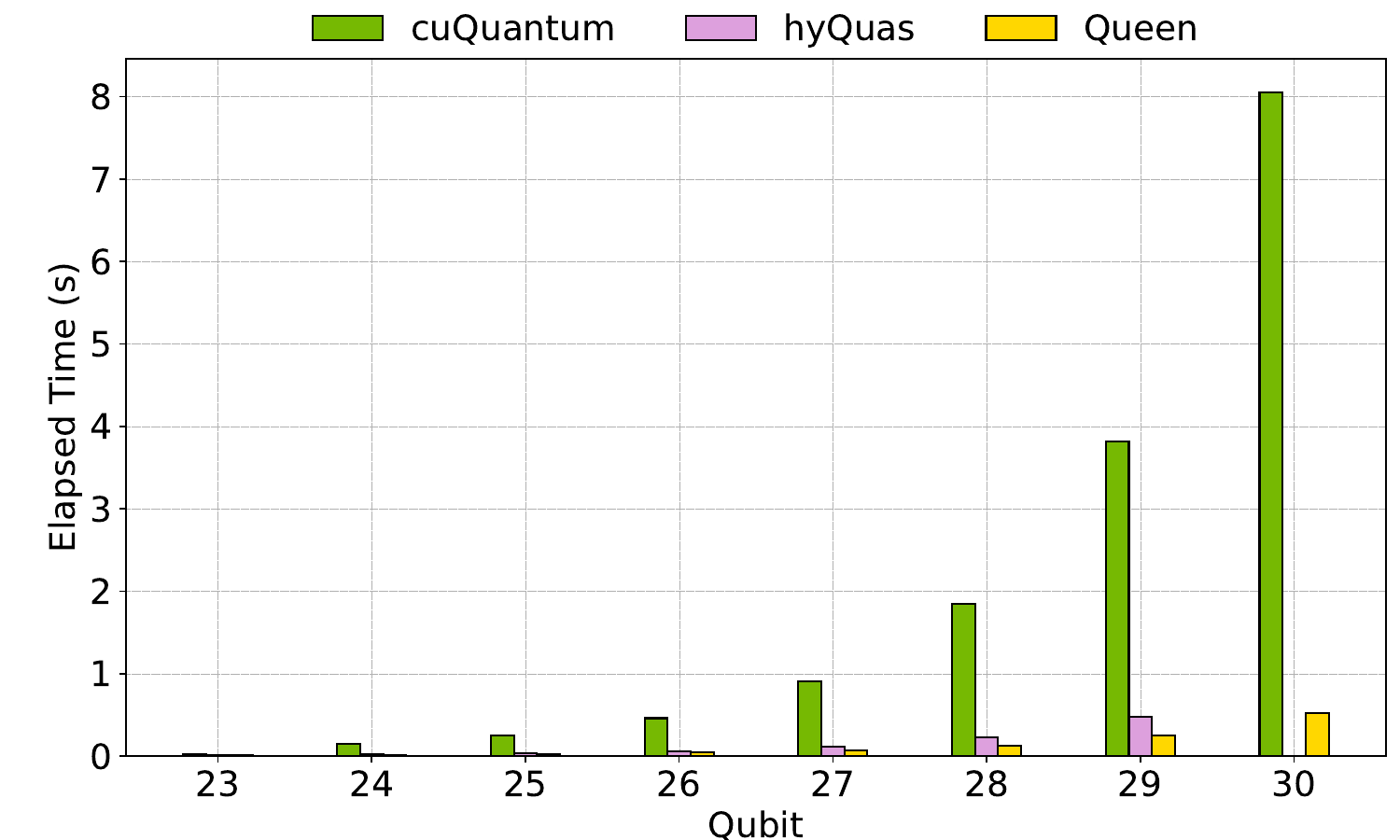}}
\caption{The QFT benchmarks with gate fusion on RTX-4090 (The time of 0 seconds for hyQuas indicates out-of-memory).}
\label{fig:qtf}
\end{figure}

\begin{figure}[tb!]
\centerline{\includegraphics[width=.95\columnwidth]{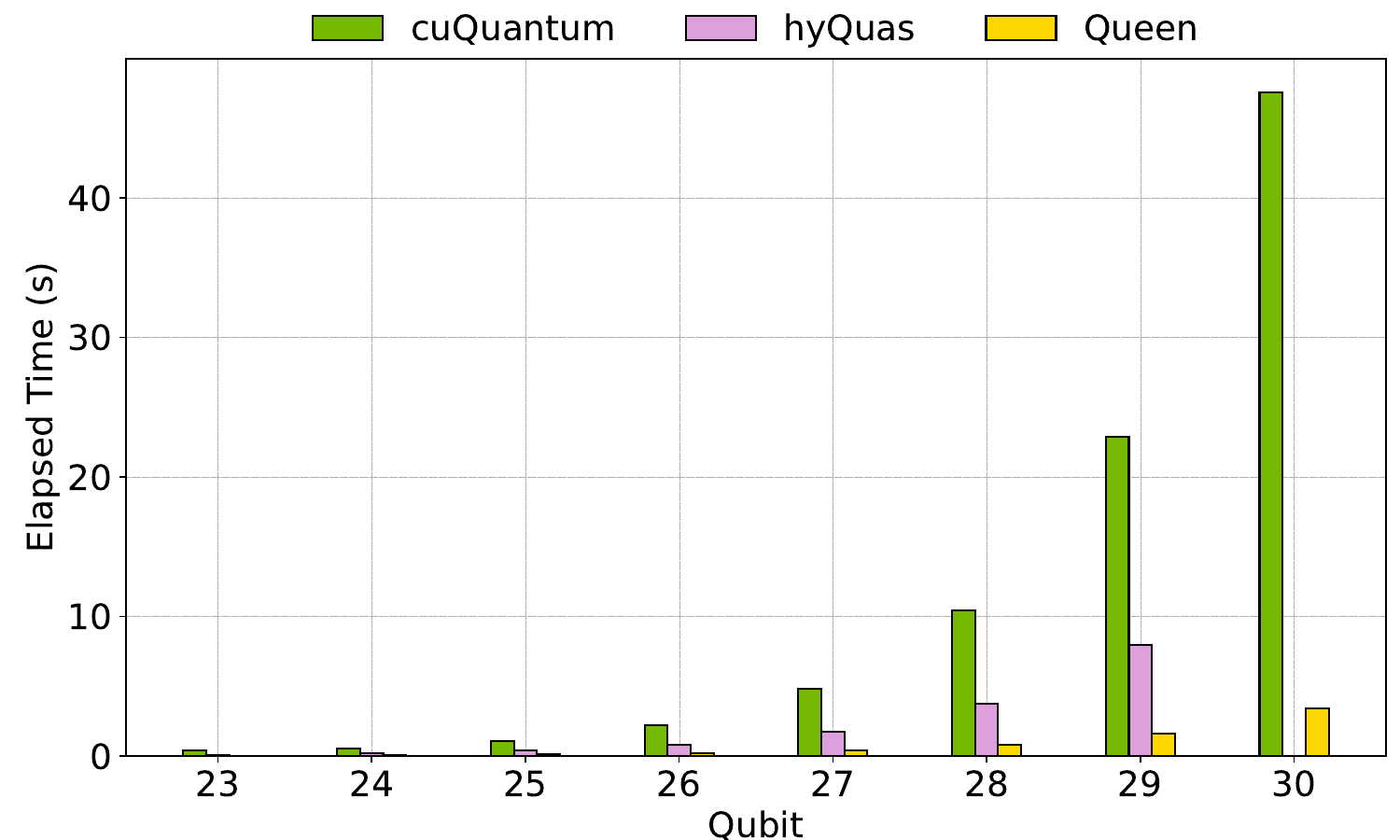}}
\caption{The QAOA benchmarks with gate fusion on RTX-4090 (The time of 0 seconds for hyQuas indicates out-of-memory).}
\label{fig:qaoa}
\end{figure}

The benchmark results are summarized in Figure~\ref{fig:gate_benchmark}, Figure~\ref{fig:qtf}, and Figure~\ref{fig:qaoa}. 
An elapsed time of 0 seconds for hyQuas indicates an out-of-memory condition.
Overall, there is still a need for additional performance analysis tools to evaluate various quantum circuit scenarios, allowing state vector-based simulators to fully realize their potential speed in high-performance computing architecture.

\section{Conclusion}
\label{sec:conclusion}
During the development cycle, there was a lot of constructive feedback to advance this work.
Appreciation is extended for applying the highest standards in reviewing this work.
It remains uncertain whether this work will be recognized as a SOTA simulator since many powerful packages and tools may not have been included in the discussion.

In summary, there are still numerous areas for performance improvement and extended research yet to be thoroughly explored, such as distributed quantum computing~\cite{dqc}, the possibilities of hybrid simulation for tensor networks and state vectors, and more.
So, let us keep the momentum going and continue pushing forward with enthusiasm.
Who knows, maybe one day, quantum circuit simulation will be faster than an actual quantum computer.

\section{Postscript}
\label{sec:postscript}
Hi, my name is Chuan-Chi Wang, and I am truly happy to connect with you through this paper. Despite facing unfair treatment, many still encourage me to bring out more truth and love.
Moving forward, I hope QueenV3 will still rock you all.



\bibliographystyle{IEEEtranS}
\bibliography{refs}

\begin{thebibliography}{10}
\providecommand{\url}[1]{#1}
\csname url@samestyle\endcsname
\providecommand{\newblock}{\relax}
\providecommand{\bibinfo}[2]{#2}
\providecommand{\BIBentrySTDinterwordspacing}{\spaceskip=0pt\relax}
\providecommand{\BIBentryALTinterwordstretchfactor}{4}
\providecommand{\BIBentryALTinterwordspacing}{\spaceskip=\fontdimen2\font plus
\BIBentryALTinterwordstretchfactor\fontdimen3\font minus \fontdimen4\font\relax}
\providecommand{\BIBforeignlanguage}[2]{{%
\expandafter\ifx\csname l@#1\endcsname\relax
\typeout{** WARNING: IEEEtranS.bst: No hyphenation pattern has been}%
\typeout{** loaded for the language `#1'. Using the pattern for}%
\typeout{** the default language instead.}%
\else
\language=\csname l@#1\endcsname
\fi
#2}}
\providecommand{\BIBdecl}{\relax}
\BIBdecl

\bibitem{Aer_sim}
\BIBentryALTinterwordspacing
``Aer - high performance quantum circuit simulation for qiskit,'' 2024. [Online]. Available: \url{https://github.com/Qiskit/qiskit-aer}
\BIBentrySTDinterwordspacing

\bibitem{cublas}
\BIBentryALTinterwordspacing
``cublas: Basic linear algebra on nvidia gpus,'' 2024. [Online]. Available: \url{https://developer.nvidia.com/cublas}
\BIBentrySTDinterwordspacing

\bibitem{cuQuantum}
H.~Bayraktar, A.~Charara, D.~Clark, S.~Cohen, T.~Costa, Y.-L.~L. Fang, Y.~Gao, J.~Guan, J.~Gunnels, A.~Haidar, A.~Hehn, M.~Hohnerbach, M.~Jones, T.~Lubowe, D.~Lyakh, S.~Morino, P.~Springer, S.~Stanwyck, I.~Terentyev, S.~Varadhan, J.~Wong, and T.~Yamaguchi, ``cuquantum sdk: A high-performance library for accelerating quantum science,'' 2023.

\bibitem{dqc}
\BIBentryALTinterwordspacing
M.~Caleffi, M.~Amoretti, D.~Ferrari, D.~Cuomo, J.~Illiano, A.~Manzalini, and A.~S. Cacciapuoti, ``Distributed quantum computing: a survey,'' 2022. [Online]. Available: \url{https://arxiv.org/abs/2212.10609}
\BIBentrySTDinterwordspacing

\bibitem{cache_blocking}
\BIBentryALTinterwordspacing
J.~Doi and H.~Horii, ``Cache blocking technique to large scale quantum computing simulation on supercomputers,'' in \emph{2020 IEEE International Conference on Quantum Computing and Engineering (QCE)}.\hskip 1em plus 0.5em minus 0.4em\relax IEEE, Oct. 2020. [Online]. Available: \url{http://dx.doi.org/10.1109/QCE49297.2020.00035}
\BIBentrySTDinterwordspacing

\bibitem{cutt}
\BIBentryALTinterwordspacing
A.-P. Hynninen and D.~I. Lyakh, ``cutt: A high-performance tensor transpose library for cuda compatible gpus,'' 2017. [Online]. Available: \url{https://arxiv.org/abs/1705.01598}
\BIBentrySTDinterwordspacing

\bibitem{45qubit}
\BIBentryALTinterwordspacing
T.~Häner and D.~S. Steiger, ``0.5 petabyte simulation of a 45-qubit quantum circuit,'' in \emph{Proceedings of the International Conference for High Performance Computing, Networking, Storage and Analysis}, ser. SC ’17.\hskip 1em plus 0.5em minus 0.4em\relax ACM, Nov. 2017. [Online]. Available: \url{http://dx.doi.org/10.1145/3126908.3126947}
\BIBentrySTDinterwordspacing

\bibitem{mpiQulacs}
S.~Imamura, M.~Yamazaki, T.~Honda, A.~Kasagi, A.~Tabuchi, H.~Nakao, N.~Fukumoto, and K.~Nakashima, ``mpiqulacs: A distributed quantum computer simulator for a64fx-based cluster systems,'' 2022.

\bibitem{cpu_gpu_communication}
\BIBentryALTinterwordspacing
C.~Jiao, W.~Zhang, and L.~Shen, ``Communication optimizations for state-vector quantum simulator on cpu+gpu clusters,'' in \emph{Proceedings of the 52nd International Conference on Parallel Processing}, ser. ICPP '23.\hskip 1em plus 0.5em minus 0.4em\relax New York, NY, USA: Association for Computing Machinery, 2023, p. 203–212. [Online]. Available: \url{https://doi.org/10.1145/3605573.3605631}
\BIBentrySTDinterwordspacing

\bibitem{QuEST}
\BIBentryALTinterwordspacing
T.~Jones, A.~Brown, I.~Bush, and S.~C. Benjamin, ``Quest and high performance simulation of quantum computers,'' \emph{Scientific Reports}, vol.~9, no.~1, Jul. 2019. [Online]. Available: \url{http://dx.doi.org/10.1038/s41598-019-47174-9}
\BIBentrySTDinterwordspacing

\bibitem{qaoa_arxiv}
Y.-C. Lin, C.-C. Wang, C.-H. Tu, and S.-H. Hung, ``Towards optimizations of quantum circuit simulation for solving max-cut problems with qaoa,'' 2023.

\bibitem{qHiPSTER}
M.~Smelyanskiy, N.~P.~D. Sawaya, and A.~Aspuru-Guzik, ``qhipster: The quantum high performance software testing environment,'' 2016.

\bibitem{Queen}
\BIBentryALTinterwordspacing
C.-C. Wang, Y.-C. Lin, Y.-J. Wang, C.-H. Tu, and S.-H. Hung, ``Queen: A quick, scalable, and comprehensive quantum circuit simulation for supercomputing,'' 2024. [Online]. Available: \url{https://arxiv.org/abs/2406.14084}
\BIBentrySTDinterwordspacing

\bibitem{HyQuas}
\BIBentryALTinterwordspacing
C.~Zhang, Z.~Song, H.~Wang, K.~Rong, and J.~Zhai, ``Hyquas: hybrid partitioner based quantum circuit simulation system on gpu,'' in \emph{Proceedings of the 35th ACM International Conference on Supercomputing}, ser. ICS '21.\hskip 1em plus 0.5em minus 0.4em\relax New York, NY, USA: Association for Computing Machinery, 2021, p. 443–454. [Online]. Available: \url{https://doi.org/10.1145/3447818.3460357}
\BIBentrySTDinterwordspacing

\bibitem{uniq}
C.~Zhang, H.~Wang, Z.~Ma, L.~Xie, Z.~Song, and J.~Zhai, ``Uniq: A unified programming model for efficient quantum circuit simulation,'' in \emph{2022 SC22: International Conference for High Performance Computing, Networking, Storage and Analysis (SC)}.\hskip 1em plus 0.5em minus 0.4em\relax IEEE Computer Society, 2022, pp. 692--707.

\end{thebibliography}

\end{document}